# Solar Radiation profiles for a 3U CubeSat in LEO

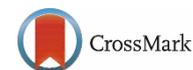

*Gianmario Broccia\*a*

[a] M.Sc. in Energy Engineering, graduated at University of Cagliari – Department of Electrical and Electronic Engineering

ABSTRACT

This paper introduces a relatively simple model to retrieve data about the solar radiation profiles on the external faces of a generic 3U-CubeSat in Low Earth Orbit (LEO). The model is inspired by relevant papers, and the related code is entirely written on Matlab. The code was intended to be well adjustable for various ranges of orbits and CubeSats with minimum effort and provide base data to be used in higher-level studies.

## 1. Introduction

The necessity of exploiting the Low Earth Orbit (LEO) for various scientific purposes, brought less specialized realities, such as Universities and even High schools, to propose interesting projects. Devised for the first time by the California Polytechnic State University (**1**), CubeSats represent the broadening of this world to a larger number of actors, also thanks to the use of commercial components for cost-saving missions. Small satellites are particularly fit to host small payloads and yet commercial components are to be considered more delicate respect space-qualified ones.

This paper comes from the necessity of creating a background about the first calculations about one of the most effective actors in LEO: sun radiation.

Here is proposed a basic model to obtain profiles of solar radiation hitting the external faces of a Cubesat of a given size along its orbit. The calculation will consider direct, reflected, and diffuse sunlight. Those values represent important boundary conditions that can be exploited in further simulations, such as multiphysics simulations in COMSOL.

The values here reported are specifically chosen for this study as a matter of example, but the model will work

\* *Corresponding author.* Tel.: (+39) 348-4665186
E-mail address: gianmario.broccia@gmail.com
Linkedin: https://www.linkedin.com/in/gianmario-broccia-827737165/

with any value, also for a sensitivity analysis.

## 2. Assumptions

A proper introduction to the model is owed and the base assumptions are briefly listed hereafter, for the sake of clarification:

1) The scenario is placed in winter, where the total radiation (from the Sun and the Earth) is at maximum **(2)**. This choice was made because the dissipation of extra heat is considered a more pressing issue than heating sensible components.

2) The attitude of the satellite will be considered with the longitudinal axis always pointed in the same direction as the tangential speed vector.

3) The faces of the satellite will be called ±X, ±Y, and ±Z according to the axis they are perpendicular to. The Z-axis will be the longitudinal one, the X will be the transversal perpendicular to the orbit plane and the Y will be parallel to the orbital plane. According to this convention, +Y sees the Earth constantly, -Y sees the outer space (external face), -Z is towards the motion direction, and +X and -X does not receive direct radiation from the sun.

4) The orbit is circular and equatorial and the satellite does not change attitude with respect to the surface, always showing the same side +Y, which might be considered the "belly" of the satellite.

## 3. Ambiental and Orbital Parameters

### 3.1 Components of incoming radiation

The first parameters to be defined are orbital and ambient. As far as concerns the direct radiation $Q_{dir}$ and the infrared radiation $Q_{IR}$ coming from the Earth, the study considers the worst-case scenario where $Q_{dir}$ is considered equal to 1414 W/m$^2$ as in the winter period **(3)**, when the distance Earth-Sun is at minimum, while the $Q_{IR}$ is quantified in 239 W/m$^2$ **(4)**, as in winter as well. The Earth's albedo $\alpha_\oplus$ is instead considered spatially uniform for the entire Earth's surface and equal to 0.3 **(5)** and will serve to consider the reflected solar radiation (part of which will be, of course, on the satellite).

### 3.2 Orbital Parameters

As far as concerns the orbital parameters, according to the Federal Aviation Administration (FAA), the LEO is characterized by an altitude equal to or less than 2400 km **(6)** so that the chosen orbital height is 500 km, which is slightly above the ISS orbit.

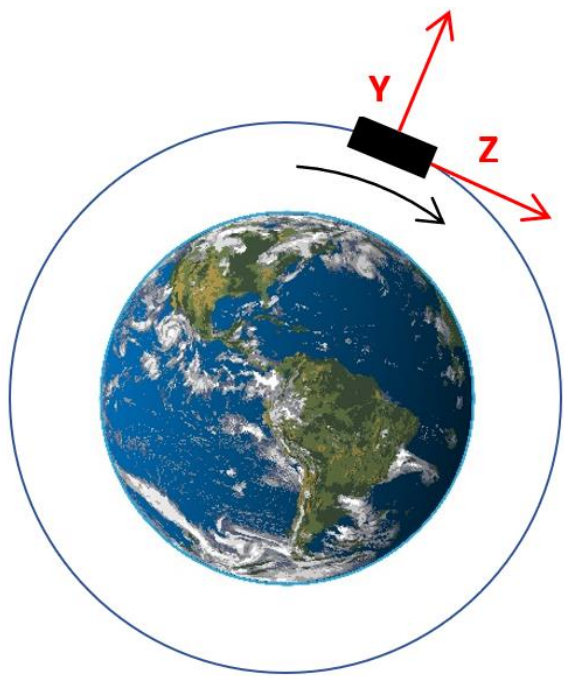

**Figure 1** - The X-axis is transversal to the orbital plane.

It must be pointed out that such parameter (as well as any other) depends on the purpose of the satellite and can be simply modified inside the code, while the whole model will work the same way. Exploiting Kepler's Third Law, an orbital period can be associated with the chosen orbital height:

$$T_{orbital} = 2\pi \sqrt{\frac{(R+z)^3}{\mu_{earth}}}$$

where:

T is the orbital period in s;

z is the orbital height (500 km);

R is the Earth's Volumetric mean Radius;

μ is the standard gravitational parameter of our planet equal to 398600.4415 km$^3$/s$^2$ **(7)**. In this case, the number of orbits considered is 30, as it showed to be the best compromise between a fast computation and a sufficiently detailed result after a few attempts. From this consideration is it clear that the simulation time will be the orbital period multiplied for the number of orbits considered.

*3.3 Shadow Zone*

It is now necessary to define the shadow zone inside which the satellite will transit. Of course, the higher the orbit is and the shorter the shadow will last. This zone will be defined by the direction the sunlight comes from which is defined by a unit vector $\hat{s}$ as in Fig.2.

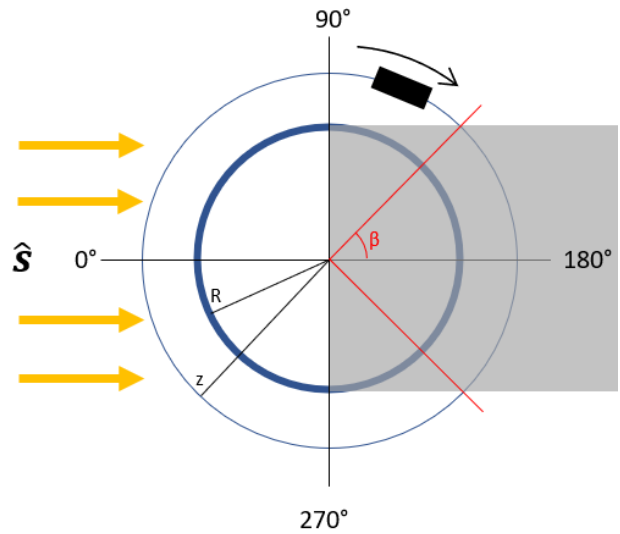

**Figure 2** - Shadow zone along the orbit.

Assuming that the zero-angle corresponds to the position in which the satellite is fully exposed to the sunlight, the shadow zone will be between 180° - β and 180° + β degrees, as in Fig.2, again. The β angle can be thus calculated as:

$$\beta = sin^{-1}(R/(R+z)) \approx 68°$$

so that the shadow zone is comprised between 112° and 248°.

## 4. Satellite Features

The chosen layout for the satellites was inspired by the ECOSAT project **(8)** and reproduced on COMSOL to deduce the size of faces, areas covered in photovoltaics, and so on. For the sake of simplification, the entire satellite is to be considered as a simple 30 x 10 x 10 cm parallelepiped which long and short faces having a total area of respectively 0.032183 and 0.0099645 $m^2$. Three faces out of six are also covered in photovoltaic cells which surface is 0.002662 $m^2$ (+Z) and 0.018634 $m^2$ (-Y and -X). An absorptivity $\alpha_{alu}$ of 0.5 for the Aluminum external faces and an $\alpha'_c$ equal to 0.95 for the solar cells was adopted. Since the solar cells have a certain efficiency $\eta_c$ (assumed equal to 0.28) that represents the amount of sunlight converted to electric energy, the real absorptivity for the cells $\alpha_c$ is:

$$\alpha_c = \alpha'_c - \eta_c = 0.67$$

Not all the faces have cells, of course, but the ones that do need a weighted average to calculate the mean value of absorptivity in that case:

$$\alpha_i = (A_{i,cells}/A_i)\alpha_{cells} + \left(1 - \left(\frac{A_{i,cells}}{A_f}\right)\right)\alpha_{alu}$$

where:

$A_i$ is the total are of the considered "i" face;

$A_{i,cells}$ is the area of the cells located in the considered face.

## 5. Angular Speed and Direct Radiation

The calculation for $Q_{dir}$ on each face is quite straightforward and only depends on the True Anomaly ν which can be defined as the angular position of the satellite with respect to the periapsis of its orbit **(9)**. This parameter is in truth undefined for circular orbits so that the zero angle is considered as in Fig.2. The true anomaly depends on the position of the satellite, known its angular speed ω expressed in rad/min:

$$\omega = 2\pi \cdot T_{orbital}$$

The position of each face respect the sun depends indeed on the angular position of the satellite along the orbit, but it must be considered that the normal vector $\hat{n}$ of each face "i" will have a different angle $\theta_i$ respect the unit vector $\hat{s}$ and the Direct Radiation on the generic face at the instant t will be:

$$Q_{dir,i}(t) = -Q_{dir}(\hat{n}_i(t) \cdot \hat{s}) \cdot \alpha_i = Q_{dir} \cdot cos\theta_i(t) \cdot \alpha_i$$

The variation of the $\theta_i$ angle can be simply represented as a harmonic motion:

$$\theta_i(t) = \omega t + j\pi = \nu(t) + j\pi$$

Where j is equal to 0 for -Y, π for +Y, π/2 for -Z, 3/2π for -Z. In the case of the X faces, the theta angle is always equal to π/2 since those are constantly perpendicular to the $\hat{s}$ unit vector and do not see the sun.

The -Y face, for example, which constantly sees the sky, will be characterized by a theta angle in phase with the True Anomaly. The face is fully exposed to the sun when the True Anomaly is zero and the cosine of the theta angle is equal to one (which corresponds to full solar radiation).

It must be also considered when the satellite enters and exits the shadow zone, of course. In the MATLAB model, this point was accounted with an IF condition that relies on cosν and senν since the transition of the satellite in certain quadrants of the orbit can be represented with a range of values for those two parameters. Again, for the sake of clarity, here is a string of the IF condition chosen for the +Y face, which sees Earth:

```
Qdir_posY = zeros(1, length(t))';

for i=1:simstep:length(t)
if (cos(w*i)>=-0.37 && cos(w*i)<0 )
Qdir_posY(i) = Qdir*cos(theta_posY(i))*alfa_posY;
else
```

```
Qdir_posY(i) = 0;
end
end
```

In this case, the minimum value of the cosine is -0.37 and refers to a value of theta of 68° or 112°, i.e. the shadow zone. The maximum value is instead zero since +Y cannot see the Sun before the True Anomaly is more than π/2 or less than 3/2π.

In case the condition is not satisfied the incident direct radiation is automatically put equal to zero since the satellite is either transiting the shadow zone or the +Y face is being shadowed by the satellite itself.

**6. Diffuse and IR Radiation and View Factors**

*6.1 View Factors*

As in the previous paragraph, the $Q_{dir}$ only depends on the inclination a given face has with respect to the Sun, which presents itself as a point-like source, but IR and reflected radiation will come from a spherical source close to the satellite and the incident radiation is not know a-priori. In this case, a proper View Factor must be considered. This parameter quantifies, in fact, the amount of radiation exiting body A that is collected on the surface of body B **(10)**.

Several relations are available in the literature and the ones of interest in this study are related to plates that receive energy from a spherical source, in the representation of the faces which are parallel or perpendicular to the Earth's surface.

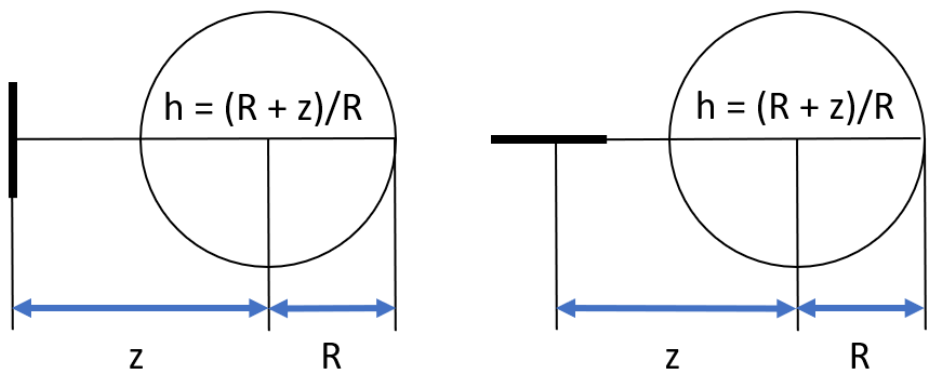

**Figure 3** - View factors for horizontal and transversal plates respect a spherical source.

These two parameters will be from now on referred respectively as $F_{ver}$ and $F_{hor}$ and can be obtained from the following relations **(11)**:

$$F_{ver} = \frac{1}{h^2} \qquad with\ h = \frac{R+z}{R}$$

$$F_{hor} = \frac{1}{\pi}\left(\left(tan^{-1}\frac{1}{x}\right) - \frac{x}{h^2}\right) \quad \text{with } x = \frac{1}{\sqrt{h^2 - 1}}$$

*6.2 IR Radiation*

The IR radiation is always present both during day and night, being independent from the presence of sunlight and it can be considered with a constant value on each side of the satellite facing the surface:

$$Q_{IR,i} = Q_{IR} \cdot \alpha_i \cdot F_{hor}$$

for the horizontal face, i.e. the ones parallel to the surface (of which +Y is the only one receiving IR radiation);

$$Q_{IR,i} = Q_{IR} \cdot \alpha_i \cdot F_{ver}$$

for the vertical faces, i.e. the ones perpendicular to the surface (+Z, -Z, +X, and -X).

*6.3 Reflected Radiation*

A different situation is for the reflected radiation coming from Earth. In this case, it is important to consider that the Earth is partially illuminated and the satellite will pass through the shadow zone.
Since the satellite transits through the shadow zone, the view factors are expected to vary along with the True Anomaly having a range between 0 (in the shadow) and $F_{ver}$ or $F_{hor}$ (in sunlight), depending on the face.
Garzon et.al (2018) **(12)** provides the results of a thorough calculation for the Libertad 2 satellite and the results are here considered.

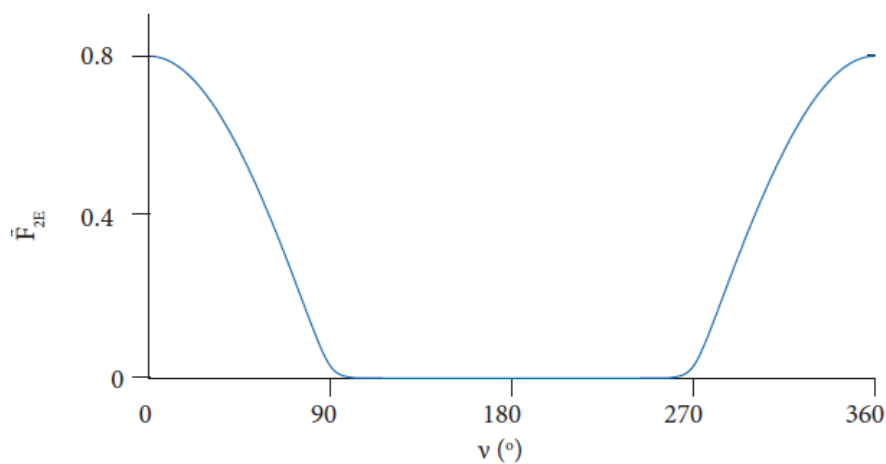

**Figure 4** - View Factor variation along the orbit for reflected radiation. Credit: (12).

In Fig. 4 the variation along the orbit for a view factor of Libertad 2's Earth-facing side is shown. Considering that Garzon does not provide analog profiles for the other faces nor cites their existence, this paper assumes that this variation in function of the True Anomaly is the same for all the faces. The only difference is supposed to be the value of the view factor, which varies between the value introduced before.

Using the MATLAB App "Curve Fitting" it was possible to obtain a polynomial function describing the variation of the View Factor. The curve was normalized to have a maximum value of 1:

$$F_{norm}(v) = 0.847 sin(0.02747v + 0.2604) + 0.9342 sin(0.04871v + 2.399) + 0.1723 sin(0.1457v + 0.9443)$$

from which the final shape of the curve is straightforward:

$$F_{hor,alb}(t) = F_{hor} F_{norm}(v(t))$$

$$F_{ver,alb}(t) = F_{ver} F_{norm}(v(t))$$

The shape of the relation describing the View Factor for the reflected radiation is thus similar to the IR radiation, but with the presence of a term dependant on the True Anomaly:

$$Q_{alb,i}(t) = (Q_{dir} \cdot \alpha_\oplus) \cdot \alpha_i \cdot F_{hor,alb}(v(t))$$

for the +Y, again, which represents the "bottom face";

$$Q_{alb,i}(t) = (Q_{dir} \cdot \alpha_\oplus) \cdot \alpha_i \cdot F_{ver,alb}(v(t))$$

for the +Z, -Z, +X, and -X.

## 7. Results

Known all the components of the radiation incident on the faces of the satellites, total profiles can be obtained from the simple summation of the relations shown before:

$$Q_i(t) = Q_{dir,i}(t) + Q_{IR,i} + Q_{alb,i}(t)$$

The resulting profile well matches what was obtained by Maria Garzon (2012) **(13)**, which also inspired a part of the code and the expected behavior in relation to the position of the satellite along the orbit. An example of one generic orbit, with the True Anomaly ranging from 0 to 360° is shown in Fig.5.

A study of this example allows to well understand the overall sunlight collection for each face. As a matter of example, the total profile on -Y only depends on the Direct Radiation and the variation matches the $\cos\theta_{negY}$, going to zero at the moment the satellite enters the shadow zone. On the other hand, +Z is the "back-face" with respect to the direction of motion, and its profile shows an increase in the collected radiation since it starts to see the sun as soon as the True Anomaly becomes greater than 0°. The total radiation has its peak at one-quarter of the orbit ($v = 90°$) the +Z fully faces the sun, going to drop to a value of $Q_{IR,posZ}\alpha_{posZ}$ when entering the shadow zone (losing both direct and reflected radiation).

It must be highlighted that the result is the actually absorbed radiation and not only the mere incident sunlight and this explains why the maximum value (for -Y) does not match the Solar Constant.

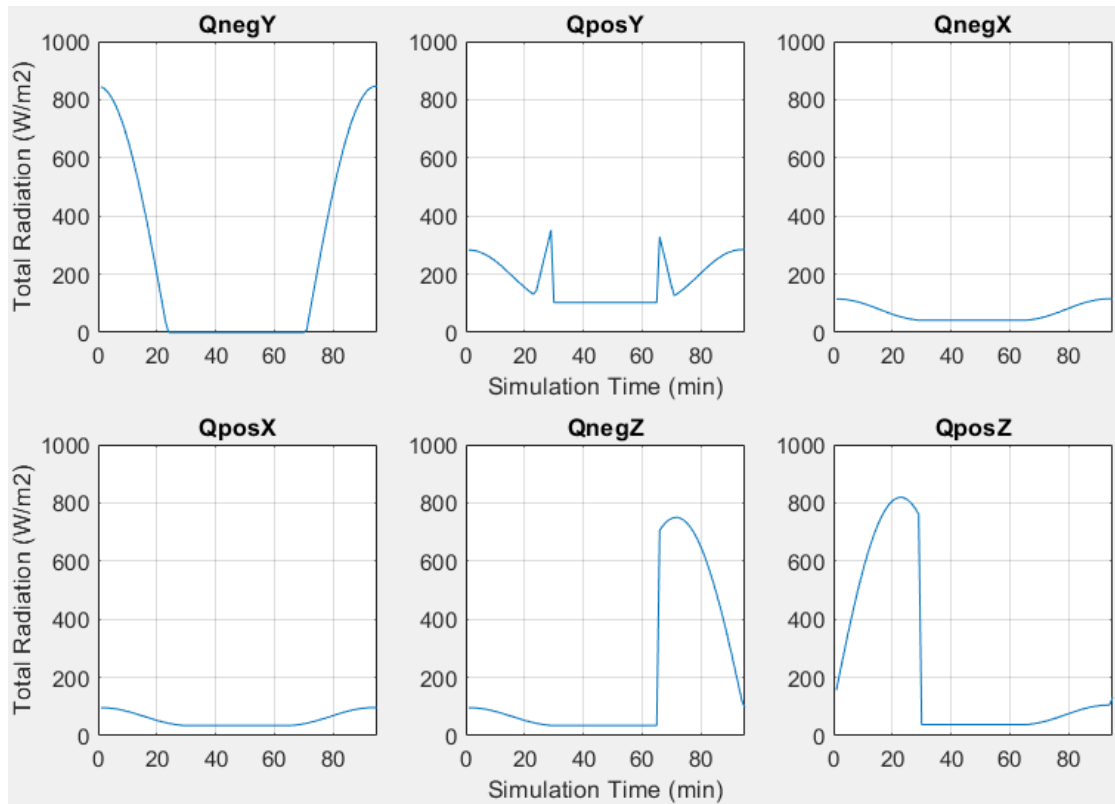

**Figure 5** - Final total profile for one generic orbit.

## 8. Conclusions

This paper showed a simple but effective model to represent the total radiation that is absorbed by a satellite along its orbit.

The model does not take into account fluctuations in the Earth's Albedo, both in time and space, and yearly variations in the Sun's activity, not to mention that the chosen orbit is the simplest possible. Yet, this model is a starting point for a fast feasibility analysis and a first step of a more complex thermal analysis for a CubeSat.

In this case, an Ecosat-like satellite was studied, showing to sensibly cut the time of a COMSOL simulation of about 50%, with respect to setting the entire model in COMSOL itself. This will allow separating the phases of the study, better identify the errors and work with a lighter simulation, using a versatile model liable of many modifications without effort.

**REFERENCES**

1. **J. Puig-Suari, C. Turner and W. Ahlgren**. *Development of the standard CubeSat deployer and a CubeSat class PicoSatellite.* California Polytechnic State University. 2001. p. 347-353. IEEE Aerospace Conference Proceedings (Cat. No.01TH8542).

2. **Cocco Daniele, Puddu Pierpaolo, Palomba chiara**. *Tecnologie delle energie rinnovabili.* s.l. : SGE snc di Modesti F. & C, 2016. p. 416. ISBN: 9788889884331.

3. **S. Corpino, M. Caldera, F. Nichele, M. Masoero, N. Viola.** *Thermal design and analysis of a nanosatellite in low earth orbit.* Politecnico di Torino & ENEA. 2015. p. 247-261. Acta Astronautica, 115.

4. **Bethel Afework, Chris Colose, Jordan Hanania, Kailyn Stenhouse, Jason Donev**. *Earth's heat balance.* University of Calgary. Last updated: May 18, 2018. https://energyeducation.ca/encyclopedia/Earth%27s_heat_balance#cite_note-3.

5. **Williams, David R**. *Earth's Fact Sheet.* NASA Goddard Space Flight Center. Last Updated: 25 November 2020. https://nssdc.gsfc.nasa.gov/planetary/factsheet/earthfact.html.

6. **FAA Website**. [Online] Last Update October 19, 2020. https://www.faa.gov/space/additional_information/faq/#s1.

7. **Watkins, J. C. Ries R. J. Eanes C. K. Shum M. M.** *Progress in the determination of the gravitational coefficient of the Earth.* 1992. p. 529-531. Geophysical research letters, 19.6 .

8. **CURRAN, Justin Thomas**. *Design and optimization of the ECOSat satellite requirements and integration: a trade study analysis of vibrational, thermal, and integration constraints .* University of Victoria, Department of Mechanical Engineering. 2014. URI: http://hdl.handle.net/1828/5843.

9. **Jack J. Lissauer, Carl D. Murray.** *Chapter 3 - Solar System Dynamics: Regular and Chaotic Motion.* s.l. : Lucy-Ann McFadden, Paul R. Weissman, Torrence V. Johnson, 2014. p. 55-79, Encyclopedia of the Solar System (Third Edition).

10. **Cengel, Yunus A**. *Heat Transfer: A practical approach 2nd Edition.* s.l. : McGraw-Hill Education , 2002. p. 960. ISBN-10 : 9780072458930 .

11. **HOWELL, John R**. *A catalog of radiation heat transfer configuration factors.* DEPARTMENT OF MECHANICAL ENGINEERING, University of Texas at Austin. s.l. : McGraw-Hill, 1982. Online Summary at: http://www.thermalradiation.net/tablecon.html.

12. **GARZÓN, Alejandro e VILLANUEVA, Yovani A.** *Thermal analysis of satellite libertad 2: a guide to cubesat temperature prediction.* Universidad Sergio Arboleda – School of Exact Sciences and Engineering – Department of Mathematics. 2018. Journal of Aerospace Technology and Management.

13. **GARZON, Maria M.** *Development and Analysis of the Thermal Design for the OSIRIS-3U CubeSat.* Pennsylvania State University, College of Engineering. 2012. Master's Thesis.